\documentclass[a4paper,twocolumn]{revtex4}
\usepackage{pdfpages}
\usepackage{bm,amsmath}
\usepackage{mathtools}
\usepackage{graphicx}
\usepackage{xcolor}
\usepackage{hyperref}
\hypersetup{
    colorlinks,
    linkcolor={red!90!black},
    citecolor={blue!90!red},
    urlcolor={blue!100!black}
}

\usepackage{mathtools}

\usepackage{psfrag}

\newcommand{\p}{\partial}
\newcommand{\f}{\frac}

\renewcommand{\d}{\mathrm{d}}

\renewcommand{\o}{\omega}

    \DeclareMathOperator{\sech}{sech}
    \DeclareMathOperator{\csch}{cosech}
   \newcommand{\dpsi}{\delta\psi}
\newcommand{\bb}{\bar{B}}
\newcommand{\bdpsi}{\delta \bar{\psi}}

\begin{document}

\title{Protein gradients in single cells induced by their coupling to ``morphogen"-like diffusion}
\author{Saroj Kumar Nandi}
\email{saroj.nandi@weizmann.ac.il}
\affiliation{Department of Chemical and Biological Physics, Weizmann Institute of Science, Rehovot 7610001, Israel}
\author{Sam A. Safran}
\email{sam.safran@weizmann.ac.il}
\affiliation{Department of Chemical and Biological Physics, Weizmann Institute of Science, Rehovot 7610001, Israel}

\begin{abstract}
One of the many ways cells transmit information within their volume is through steady spatial gradients of different proteins. However, the mechanism through which proteins without any sources or sinks form such single-cell gradients is not yet fully understood. One of the models for such gradient formation, based on differential diffusion, is limited to proteins with large ratios of their diffusion constants or to specific protein-large molecule interactions.
We introduce a novel mechanism for gradient formation via the coupling of the proteins within a single cell with a molecule, that we call a ``pronogen'', whose action is similar to that of morphogens in multi-cell assemblies; the pronogen is produced with a fixed flux at one side of the cell. This coupling results in an effectively non-linear diffusion degradation model for the pronogen dynamics within the cell, which leads to a steady-state gradient of the protein concentration. We use a stability analysis to show that these gradients are linearly stable with respect to perturbations.
\end{abstract}

\maketitle

\section{Introduction}
The spatial regulation of different active processes at various stages of development presents an intriguing problem of biological physics. A large number of recent observations suggest that intracellular as well as extracellular gradients of various biological molecules are one means of imparting spatial information over a range of length scales for different processes \cite{fuller2010,kiekebusch2012,wartlick2009,gorlich2003,caudron2005,kalab2006,tostevin2011}. The initiation of mitosis in fission yeast \cite{fantes1977,nurse1977,moseley2009,tostevin2011}, chromosome organization and DNA replication in bacteria \cite{shapiro2009,tropini2012}, chemotaxis in {\it Escherichia coli} \cite{vaknin2004,lipkow2005}, germline formation in {\it C. elegans} \cite{daniels2009,samba2016} are some examples.
It has also been shown that such gradients of morphogens or proteins are quite robust to perturbations \cite{eldar2002,eldar2003} and in some cases, may scale with the system size \cite{benzvi2008,benzvi2009,benzvi2011,dasbiswas2016}.

These biological examples have motivated theoretical interest in the dynamics that can lead to nonuniform, steady-state diffusion-like profiles.  In this paper, we focus on the generic physical aspects of such models that are more general than those constructed to explain a particular biological experiment. An early model by Crick shows that freely diffusing morphogen produced at a source and degraded at a sink at a distant point produces a ``linear'' gradient at a biologically relevant time scale \cite{crick1970}. It is by now well-known that a ``localized'' sink is not necessary for gradient formation; uniform degradation throughout the system can also result in molecular gradients \cite{wartlick2009,kicheva2007}. 
Recently suggested mechanisms for gradient formation include protein binding to the cell membrane, via the recognition of membrane curvature \cite{ramamurthi2010} or gradients formed via protein-protein interactions \cite{wettmann2014}.
In the case of some proteins, gradients can be formed by localized protein sources and sinks separated in space; the gradient formation of Bicoid proteins in Drosophila embryos \cite{driever1988a,driever1988b,little2011,grimm2010} and that of Fgf8 in living zebrafish \cite{yu2009} constitute two such examples . 

However, there are other types of proteins that form gradients without any cellular source or sink of the proteins \cite{brown1999,howard2012,griffin2011,hachet2011,saunders2012,hersch2015,samba2016}; the mechanism behind such gradients is not yet well-understood. 
Apart from some kinetic modeling \cite{caudron2005,gorlich2003}, 
most of our understanding for this problem comes from the differential diffusion model 
of two or more states (e.g., phosphorylation states) of the same protein that, for example, are introduced by a localized source of phosphorylation (addition of phosphate group ($PO_4^{3-}$)) and spatially uniform dephosphorylation \cite{lipkow2008,daniels2009,daniels2010}. 
This can be generalized to other chemical modification of molecules. These diffusion models are crucially dependent on there being a relatively large difference in diffusivity of the protein upon phosphorylation [see Supplementary Material (SM)].
They are therefore limited to specific protein-large molecule associations, regulated by phosphorylation, that lead to an effectively much larger protein complex, since phosphorylation by itself does not significantly change the diffusivity of proteins \cite{lipkow2008,koffman2015}. Moreover, as shown in the SM, the mere presence of slow and fast moving species of the protein does not necessarily imply the applicability of the differential diffusion mechanism. 
This motivates the need for alternative mechanisms such as the one presented in this paper, which is based on the interactions of the proteins with gradients of molecules whose action within a single cell is similar to that of morphogens in cell assemblies.

We show in this paper that proteins, without any sources or sinks, may form gradients if the proteins are coupled to other molecules that we denote as ``pronogens'' that act within a single-cell. 
Pronogens are defined as small diffusing molecules, whose action is similar to that of morphogens in multi-cellular assemblies. The pronogens are secreted at one end of the cell and the gradients they form template the concentration profile of proteins with which they interact
A localized source for the pronogens along with their degradation that occurs throughout the cell, can result in a steady-state concentration gradient of the pronogen \cite{wartlick2009,teimouri2016}. Proteins that associate with such pronogen may then also develop a concentration gradient. It is already well-known that some kind of pronogen are involved in the gradient formation of proteins that do not have any sources or sinks. 
Examples include, the gradient of nuclear localization signal (NLS)-containing proteins during spatial coordination of the spindle assembly~\cite{gorlich2003,caudron2005} and gradient formation of MEX-5 proteins at one-cell stage of {\it C. elegans} embryo \cite{samba2016}. While our focus is theoretical, a discussion of the relevance of the various models to experiment is presented in the Discussion section, including suggestions for future experiments.

We consider a quasi-static scenario where the pronogen dynamics, with a localized source and uniform degradation, is treated within a nonequilibrium framework whereas the protein dynamics is taken to have reached equilibrium. We outline the conditions where this latter approximation is relevant in the SM. We show that there are two interesting scenarios depending on the sign of the protein-pronogen interaction; the protein profile may follow or be complementary to that of the pronogen profile. 

The rest of the paper is organized as follows: for protein gradient formation, where a diffusing pronogen produced at a localized source, is uniformly removed from the solution (effectively degraded) through its association with the protein of interest. In general, this association can either lower or raise the free energy of the protein.
Pronogen ``degradation'' and diffusion results in a steady-state pronogen concentration profile within the single cell.
The association of the pronogen with the more slowly diffusing protein then results in a protein gradient within the single cell. The model may be applicable to a wide variety of systems. The nature of coupling of the protein-pronogen interaction leads to two different scenarios whose solutions are respectively presented in Secs. \ref{solution_model1} and \ref{solution_model2}. We discuss the consequences of our model and its predictions in Sec. \ref{discussion}.

\section{Gradient formation through pronogen diffusion}
\label{ourmodel}
We first present our model that shows how a pronogen (that diffuses within a single cell) can interact with proteins to form both pronogen and protein steady-state gradients in biological systems. A similar idea coupling one steady-state, reaction-determined (morphogen) concentration profile to another species that reaches equilibrium (but on the scale of cell assemblies) was first considered by Dasbiswas, Alster and Safran in the context of a mechanobiological model for the coupling of long-range contractility (that reaches mechanical equilibrium) and diffusing biomolecules to show how the morphogen concentration profile can scale with the system size due to long-range mechanical coupling \cite{dasbiswas2016}.
We present our model, where the pronogen have a steady-state profile but the proteins can reach thermodynamic -- as opposed to mechanical -- equilibrium, in general terms since the physics and predictions are generic in nature, but do discuss specific applications.
Our model predicts protein gradient formation via the simplest possible non-linear term (quadratic in the concentration) in the effective diffusion-degradation equation for the pronogen.
We note that since the proteins have no long-range coupling with each other, we do not find scaling of the protein concentration profile with the system size as was the case for the morphogen-contractility coupling in multi-cellular assemblies of Ref. \cite{dasbiswas2016}. Our focus here is on how protein gradients within single cells can be established and not on scaling with system size.

\begin{figure}
 \includegraphics[width=8.6cm]{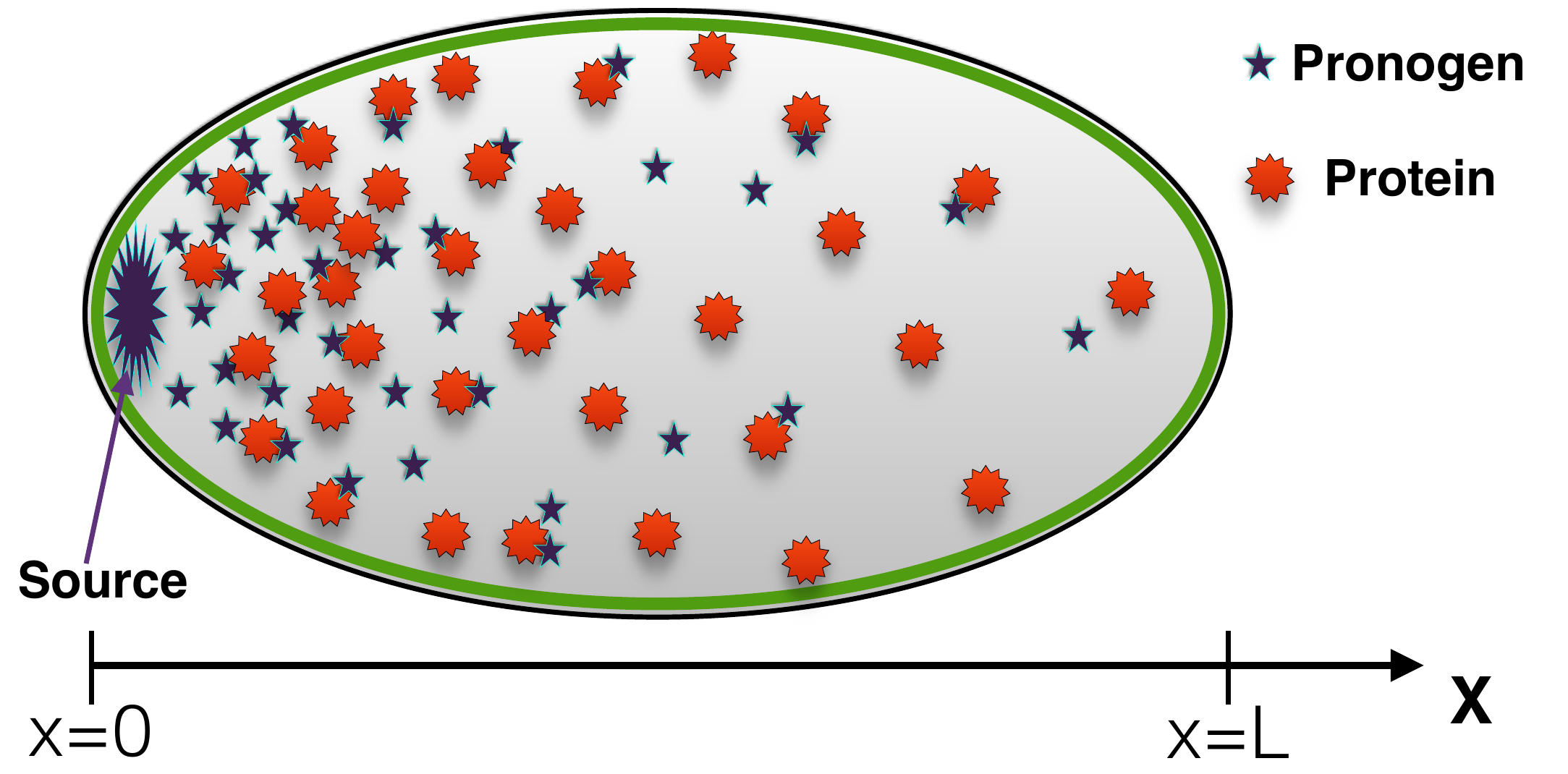}
 \caption{A schematic picture of gradient formation of protein via pronogen diffusion. The pronogen has a fixed flux at one end of the cell, at $x=0$, and is degraded by association with the protein molecules throughout the bulk. In the steady state, the pronogen concentration has a steady-state profile. 
 We predict here how the protein-pronogen interactions result in a protein concentration gradient.}
 \label{morphogen_model}
\end{figure}

In our description, the pronogen acts as a signal within a single cell that establishes a gradient of protein at a certain stage of development (see Fig. \ref{morphogen_model} for a schematic picture). Such a pronogen, for example, may correspond to a phosphate ($PO_4^{3-}$) group in case of {\it C. elegans} or to GTPase Ran in case of spindle formation in early embryo development (e.g. Xenopus). This pronogen is created at localized source that is expressed by imposing a flux-boundary condition; once secreted, the pronogen diffuses throughout the cell with diffusivity $D$. The diffusing pronogen locally associates throughout the cell volume, with the proteins of interest (assumed to be uniformly distributed before the pronogen is secreted), that eventually adjust their local concentration to that of the pronogen. This ``adiabatic'' approximation assumes that the pronogen diffuses quickly to attain its steady-state profile determined by the instantaneous protein concentration which changes slowly. The proteins eventually reach equilibrium as determined by minimization of the local protein free energy, including its interaction with the local concentration of the pronogen (see SM).
Therefore, the diffusion-degradation equation, describing the dynamics of the free pronogen concentration, $\psi(x,t)$ (in dimensionless units where the volume is scaled by the pronogen molecular volume), at position $x$ and time $t$ will be
\begin{equation}\label{phosphoeq1}
 \f{\p\psi(x,t)}{\p t}=D\nabla^2\psi(x,t)-R_0\psi(x,t)- \bar{\Lambda}[c(x,t),\psi(x,t)],
\end{equation}
where $R_0$ is the rate of local degradation that may include irreversible association of the pronogen with other molecules as well as biochemical degradation. The effective degradation term represented by $\bar{\Lambda}[c(x,t),\psi(x,t)]$ accounts for the interaction of the pronogen with the gradient-forming protein whose local concentration evolves much more slowly (than pronogen diffusion) with time. 
In the absence of coupling to the protein ($\bar  \Lambda =0$), we have  a simple diffusion-degradation equation. The steady-state balance of the localized flux of pronogen at the origin with the homogeneous degradation of the molecule in the bulk leads to an exponentially decaying pronogen concentration, where the decay length is the square root of the ratio of the two kinetic coefficients, $D$ and $R_0$.

We now focus on the new effects due to pronogen-protein ccoupling; the most interesting phenomena occur when the pronogen-protein associations are irreversible and result in pronogen degradation.
Therefore, $\bar{\Lambda}[c(x,t),\psi(x,t)]$ is proportional to the local protein concentration, $c(x,t)$. However, there must be finite amount of pronogen for degradation, therefore, $\bar{\Lambda}[c(x,t),\psi(x,t)]$ is also proportional to $\psi(x,t)$. Thus, we have $\bar{\Lambda}[c(x,t),\psi(x,t)]=\bar{A}c(x,t)\psi(x,t)$, where $\bar{A}$ is a rate constant.

The protein concentration dynamics is obtained  from a generalized diffusion equation that takes into account the protein-pronogen interactions. We assume that the protein lifetime is much longer than that of the pronogen so that it does not biochemicallly degrade during the time scale that the pronogen concentration reaches its steady state or for the time scale of the relevant experiments. This is in contrast to the pronogen whose degradation time is taken to be much smaller. In this case, the protein dynamics are related to the derivative of the protein flux, given in a first approximation, to the gradient of the local chemical potential which is not constant until the system reaches equilibrium. The local chemical potential is obtained from the local protein free energy, per unit volume $f$, that includes the local entropy and interactions with the pronogen:
\begin{align}\label{phosphoeq3}
f[c(x,t),\psi(x,t)]=c(x,t)(\ln c(x,t)-1)-\alpha c(x,t) \psi(x,t)
\end{align}
where $\alpha$ is the protein-pronogen interaction strength and we have set $k_BT$, the Boltzmann constant times temperature, to unity. The chemical potential $\mu(x,t )=\partial f/\partial c(x,t)$. Then the protein concentration dynamics is obtained as
\begin{equation}
 \f{\p c(x,t)}{\p t}=D_p\nabla\cdot[c(x,t)\nabla\mu(x,t)]=D_p\nabla[\nabla c-c\alpha\nabla\psi]
\end{equation}
where $D_p$ is the diffusivity of the protein. Then, in the steady state, we have
\begin{equation}\label{ss_protein}
 \nabla c(x,t)-c(x,t)\alpha\nabla\psi(x,t)=K
\end{equation}
where $K$ is a constant. $K$ is non-zero in general and zero in equilibrium. 
In the SM we show that in the case where the proteins do not degrade (or their degradation time is much longer than that of the pronogen of the relevant experiments) that they reach a local equilibrium with the pronogen concentration and hence $K=0$. In that case, we can solve Eq. (\ref{ss_protein}) to obtain:
\begin{equation}\label{phosphoeq4}
c(x,t)=c_0 \exp[\alpha \psi(x,t)]
\end{equation}
where $c_0$ is a constant which in equilibrium, is $c_0=\exp[\mu]$. 
Thus, at long times, the proteins reach local equilibrium via their interactions with the steady-state pronogen profile (see SM) and we solve Eq. (\ref{phosphoeq1}) in that limit where the protein concentration profile in equilibrium is denoted by $c(x)$.

In the cases we consider, the pronogen can be a phosphate group or an enzyme that diffuses quickly.  The protein concentration profile is assumed to be uniform before the pronogen is secreted at one end of the cell (this can also be somwhere in the bulk of the cell, as in the case of spindle formation). The pronogen quickly reaches its steady-state profile. The proteins, which diffuse more slowly, interact with this steady-state profile and then locally diffuse to adjust their own concentration profile to adjust to the local pronogen concentration.
As shown above and in more detail in the SM, the proteins eventually reach their local, equilibrium concentration profile, $c(x)$, that includes their interactions with the pronogen. Here we allow the protein-pronogen interaction coefficient $\alpha$ to be either $+$ or $-$. Positive $\alpha$ decreases the protein free energy and such processes are always allowed. Negative values of $\alpha$ increase the protein free energy and such processes are rare, but still allowed in equilibrium due to entropic effects. Biological systems are generally out of equilibrium, so that processes that can increase the free energy are still allowed. In this case, the pronogen will antagonize the protein. The two signs of the interaction strength $\alpha$ lead to two different situations that we solve separately. We call the situation where $\alpha>0$, the synergistic protein-pronogen interaction and the situation where $\alpha<0$, the antagonistic protein-pronogen interaction, hereafter denoted as synergistic and antagonistic respectively.

We now use the solution for the local equilibrium protein concentration:  $c(x) = \exp[\mu] \exp[\alpha \psi(x,t)]$ in the equation for the pronogen diffusion to obtain self-consistent expressions for the pronogen and protein concentrations.
We can, in principle, solve for any values of $\alpha$, but it is more instructive (but only for mathematical convenience) to consider the small interaction case, which is the focus in this paper . In this limit we expand the exponential and obtain $c(x)=e^\mu(1+\alpha\psi)$. Using this, we obtain from Eq. (\ref{phosphoeq1}), for the synergistic scenario
\begin{equation}\label{phosphoeq5}
\f{\p\psi(x,t)}{\p t}=D\nabla^2\psi(x,t)- A\psi(x,t) - B\psi^2(x,t),
\end{equation}
where $A=R_0+\bar{A}e^\mu$ and $B=\bar{A}e^\mu\alpha$. For the antagonistic scenario, when $\alpha<0$, we obtain
\begin{equation}\label{phosphoeq6}
\f{\p\psi(x,t)}{\p t}=D\nabla^2\psi(x,t)- A\psi(x,t) + B\psi^2(x,t),
\end{equation}
with $B=\bar{A}e^\mu|\alpha|$.
We see from Eq. (\ref{phosphoeq4}) that the protein concentration profile follows the pronogen concentration for the synergistic interaction, whereas the antagonistic scenario predicts a protein concentration profile that is complementary to that of the pronogen concentration.

When $\alpha\approx 0$, $e^\mu$ is approximately the average protein concentration $c_0$. 
For non-zero $\alpha$, $\mu\approx \ln c_0+\mathcal{O}(1/L)$ where $L$ is the $1d$ size of the system. Considering $L$ very large with respect to the molecular interactions that govern the pronogen concentration profile, we obtain $c_0\approx e^\mu$ and therefore
\begin{equation}\label{cxc0}
 c(x)=c_0e^{\alpha \psi}
\end{equation}
We set the time-derivatives of the pronogen concentration $\psi$ in Eqs. (\ref{phosphoeq5}) and (\ref{phosphoeq6}) to zero in the steady-state and also consider the equilibrium protein concentration profile. To complete the description of the models, we must account for the two boundary conditions on $\psi(x)$. Since the pronogen is secreted at the $x=0$ edge of the cell with a fixed flux, one boundary condition is $\d\psi(x,t)/\d x=-j_0$ where $j_0$ is the pronogen flux. The other boundary condition is that the pronogen flux vanishes at the other end of the cell, thus, $\d\psi(x,t)/\d x=0$. For simplicity, we take the cell size $L\to\infty$ (relative to molecular scales); however, this doesn't affect the results if $L$ is much greater than the typical molecular scales where the concentration gradients are significant.


\section{Synergistic protein-pronogen interactions}
\label{solution_model1}
We first consider the synergistic situation where the protein energy is  decreased by association with the pronogen ($\alpha>0$), with the equation of motion for the pronogen density in an infinite, one-dimensional system given by:
\begin{equation}\label{synergmodel}
\f{\p\psi(x,t)}{\p t}=D\nabla^2\psi(x,t)- A\psi(x,t) - B\psi^2(x,t),
\end{equation}
with the boundary condition of a fixed flux $j_0$ at $x=0$ and zero flux at $x=L$, where we focus on the limit where $L \rightarrow \infty$, relative to the molecular length scales of the gradients. We write the steady state solution of $\psi(x,t)=\phi(x)$ with 
\begin{equation}
\f{\d\phi(x)}{\d x}\bigg|_{x=0}=-j_0,
\end{equation}
and show in the SM that this steady-state solution is stable to small perturbations of the profile. To simplify the discussion, we define $\phi_0=\phi/j_0$ and $\bb=j_0B/A$ and redefine $x\to \sqrt{A}x$ and $t\to A t$. We also measure $j_0$ in units of $\sqrt{A}$ and set $D$ to unity. (Note that these redefinitions do not affect the nature of the solution; they can be reversed at the end to give dimensional times, distances and fluxes.)
After a first integration and taking $L\to\infty$, we obtain (see SM for details)
\begin{equation}
 \f{\d\phi_0}{\d x}=\pm \phi_0\left(1+\f{2\bb}{3}\phi_0\right)^{1/2}.
\end{equation}

Taking the $+$ sign ($-$ sign leads to the same results) and integrating once more we obtain the steady-state pronogen profile:
\begin{equation}\label{sol_model1_eq20}
 \phi_0(x)=\f{3}{2\bb}\text{cosech}^2\left[\f{x}{2}+k\right]
\end{equation}
where $k$ is an integration constant determined from the boundary condition. With the previous normalizations, we can now write the boundary condition at $x=0$ as $\d\phi_0/\d x|_{x=0}=-1$, so that
\begin{equation}\label{bc_model1_eq15}
 \coth k(\coth^2k-1)=\f{2\bb}{3}.
\end{equation}
The parameter $\bb$ that determines $k$ is proportional to the product of the pronogen flux, $j_0$, and the pronogen-protein interaction, $\alpha$.

\begin{figure}
 \includegraphics[width=8.6cm]{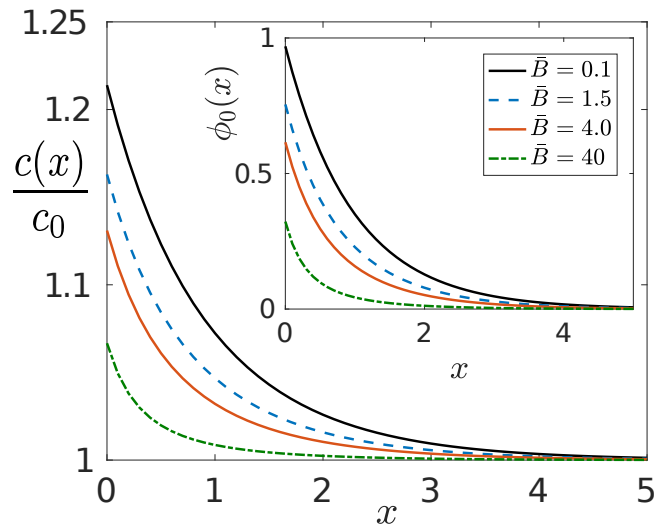}
 \caption{We plot the equilibrium protein profile $c(x)/c_0$, Eq. (\ref{cxc0}), for an interaction (in units of $k_BT$) $\alpha=0.2$ and different values of $\bb$ as shown in the figure. For non-zero $\alpha$, $c_0=\ln\mu +\mathcal{O}(1/L)$ is constant and $c(x)/c_0$ determines the profile of the total protein concentration. {\bf Inset:} The pronogen profile for the same parameters with a source of pronogen at $x=0$. Note that we rescaled the spatial coordinate by $1/\sqrt{A}$ where $A$ is the coefficient of the linear term [Eq. (\ref{synergmodel})].}
 \label{model1_phi_psi}
\end{figure}

Eq. (\ref{bc_model1_eq15}) has only one real solution for $k$ and in the SM we perform a stability analysis to show this solution is stable.
Once we obtain the solution for $\phi_0(x)$, which is the steady state solution for $\psi(x,t)$, we can find the local concentration of the protein from Eq. (\ref{phosphoeq4}). Since $c_0\approx \ln \mu$ with a correction of the order of $1/L$ is a constant, $c(x)/c_0$ determines the profile of the total protein concentration. We show the profiles for 
$c(x)/c_0$ for different values of $\bb$ in Fig. \ref{model1_phi_psi} and the corresponding profiles for the pronogen concentration $\phi_0(x)$ in the inset.
Note that we have rescaled the spatial coordinate $x$ by $1/\sqrt{A}$. 

The curves go to zero more slowly as the parameter $\bb$ decreases. Thus, the cell can reduce $\bb$ (through reducing the flux, for example) to keep the protein concentration significant over larger length scales. On the other hand, larger values of $\bb$ result in sharper spatial decay of the protein concentration. Thus, $\bb$ (which includes both the protein-pronogen interaction and the flux in its definition) is a single parameter that controls the protein concentration gradient.

\section{Antagonistic protein-pronogen interactions}
\label{solution_model2}
We now consider the antagonistic situation where the protein energy increases by association with the pronogen ($\alpha<0$), where the equation of motion for the pronogen density is governed by
\begin{equation}\label{model2_eq26}
\f{\p\psi(x,t)}{\p t}=D\nabla^2\psi(x,t)- A\psi(x,t) + B\psi^2(x,t),
\end{equation}
with the boundary condition of a fixed flux $j_0$ at $x=0$ while the flux vanishes at $\infty$. Scaling $\phi(x)$ by $j_0$ and measuring $j_0$ in units of $\sqrt{A}$, we write the boundary condition at $x=0$ as $\d\phi_0(x)/\d x|_{x=0}=-1$, where $\phi_0(x)=\phi(x)/j_0$. Setting $D$ to unity and scaling $x\to\sqrt{A}x$ and $t\to At$ as before, we obtain the solution for the steady state pronogen profile as (see SM for more detail)
\begin{equation}
 \phi_0(x)=\f{3}{2\bb}\sech^2\left[\f{x}{2}+k\right],
\end{equation}
where $k$ is the integration constant that is determined through the boundary condition at $x=0$:
\begin{equation}\label{model2_eq31}
 \tanh k(1-\tanh^2k)=\f{2\bb}{3}.
\end{equation}
\begin{figure}
 \includegraphics[width=8.6cm]{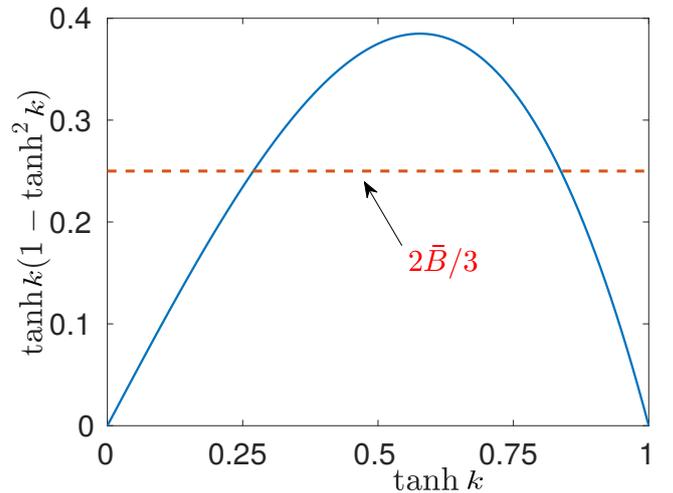}
 \caption{The graphical solution for the undetermined constant, $k$, obtained from the boundary condition, Eq. (\ref{model2_eq31}), shows that there can be two real solutions for $k$. One solution is stable and the other is unstable that evolves in time to the stable solution if slightly perturbed.}
 \label{model2_fluxbc}
\end{figure}
Since $\tanh k$ ranges from 0 to 1, Eq. (\ref{model2_eq31}) imposes a constraint on $\bb$. Using the maximum of the left hand side of Eq. (\ref{model2_eq31}), we find that in order to have a real solution for the integration constant $k$, the parameter $\bb\leq 1/\sqrt{3}$. If $\bb$ is greater than this value, there is no steady-state solution. From Fig. \ref{model2_fluxbc} we see that for certain values of $\bb$, we have two real solutions for $k$. From the stability analysis, we find that the solution corresponding to one of these values of $k$ is stable, whereas the other is unstable and dynamically evolves when perturbed to the stable solution. If $\bb\phi_0(x=0)$ is greater than unity, the solution is unstable; otherwise it is stable (see SM). 
The third solution of Eq. (\ref{model2_eq31}) gives a complex value for $k$ that leads to negative values for $\phi_0(x)$. Since $\phi_0(x)$ is a concentration, it must be positive so that we do not consider this as a physical solution.

\begin{figure}
 \includegraphics[width=8.6cm]{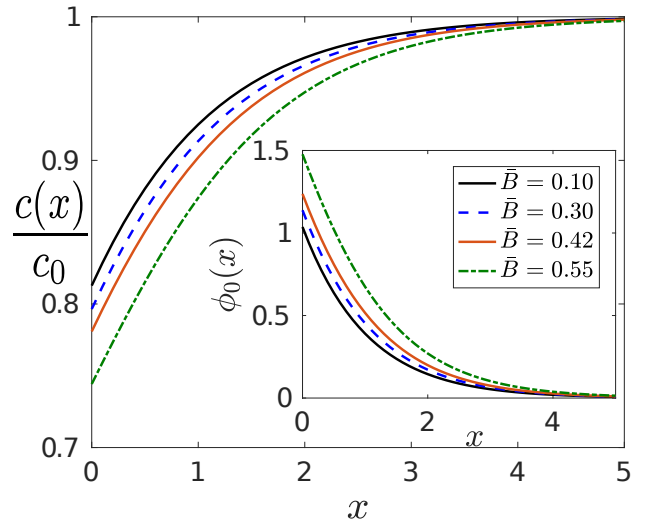}
 \caption{Plot of the stable solution for the protein concentration profile, scaled by $c_0\approx\ln\mu$, for different values of $\bb$ with the interaction, $\alpha=-0.2$. {\bf Inset:} Solution for the pronogen concentration for the corresponding values of the parameters. We see that the gradient of protein concentration is complementary to that of the pronogen concentration, which is expected since the pronogen increases the protein energy in this antagonistic case, so the protein avoids the pronogen.}
 \label{model2_psi_cu}
\end{figure}

We plot the protein concentration profile $c(x)/c_0$, Eq. (\ref{cxc0}), in Fig. \ref{model2_psi_cu} for different values of $\bb$ as shown in the figure and the pronogen concentration for the corresponding parameters in the inset. Since $\alpha=-0.2$ is negative, in this antagonistic case, the protein concentration profile is complementary to that of the pronogen profile; the proteins tend to avoid the pronogen. The phosphorylation process of MEX-5 protein in the one-cell stage of {\it C. elegans} can be taken as an example of this case. In the unphosphorylated state, MEX-5 proteins associate with larger molecules through RNA. When phosphorylated, MEX-5 dissociates from RNA leading to a local entropy increase; the proteins thus tend to avoid the phosphate molecules.


\section{Discussion} 
\label{discussion}
We have shown that protein gradients in single cells can be established from an initially, uniform protein distribution via their interactions with a pronogen, which is a diffusing molecule whose action is similar to that of morphogen produced at a localized source; both the protein and pronogen dynamics are coupled. 
Rescaling the steady state pronogen concentration by its flux allows us to predict the concentration profiles as a function of a single parameter $\bar{B}=j_0B/A$ where $j_0$ is the pronogen flux at one end of the cell. The parameters $B$ and $A$ respectively account for the quadratic and linear degradations.
This parametrization -- related to its flux -- is advantageous since the gradient of the pronogen (and the protein gradient)
are the important parameters rather than the values of the concentration at the source, located at $x=0$.
This permits the cell to regulate the protein gradient via changes in the pronogen flux.
For example, Figs. \ref{model1_phi_psi} and \ref{model2_psi_cu} show that the profiles for the pronogen and the protein strongly depend on $\bb$ and cells can adjust this parameter in order to realize different profiles for the protein concentration.
Examples of biological systems, where our theory should be relevant, include the gradient formation of nuclear localization signal (NLS)-containing proteins, influenced by the gradient of GTPase (guanosine triphosphatase) Ran, during spatial coordination of the spindle assembly~\cite{gorlich2003,caudron2005} and gradient formation of MEX-5 proteins, influenced through localized phosphorylation at the posterior side, inside the {\it C. elegans} embryo at one-cell stage \cite{samba2016}.

\begin{figure}
\includegraphics[width=8.6cm]{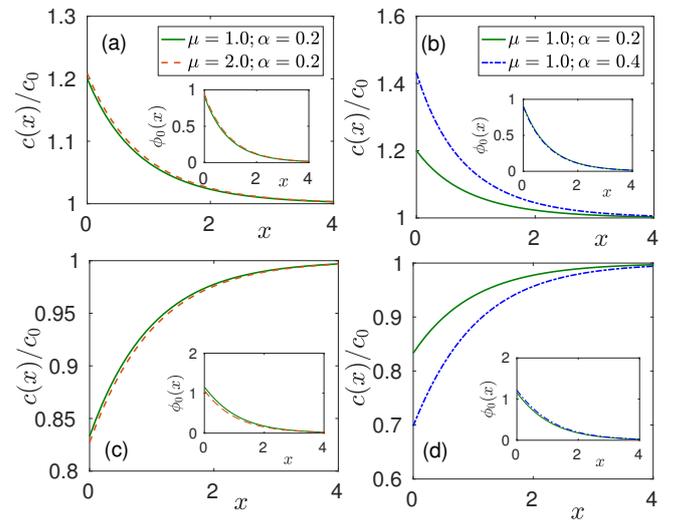}
 \caption{Effects on the protein gradient with changes in the protein-pronogen interaction strength and protein chemical potential. (a) and (b) are for the synergistic scenario and (c) and (d) are for the antagonistic scenario. (a) The protein gradient changes slightly when the protein chemical potential $\mu$ changes from $1$ to $2$ for fixed protein-pronogen interaction strength $\alpha=0.2$. (b) The protein gradient is significantly affected when $\alpha$ changes from 0.2 to 0.4 at fixed $\mu=1.0$. (c) and (d) shows the effects on the protein gradient for the antagonistic scenario (negative values of $\alpha$) for the changes in parameters used in (a) and (b) respectively.}
 \label{diffmu_alpha}
\end{figure}

We now discuss several ways to test our theory by changing the experimental parameters as shown in Fig. \ref{diffmu_alpha}. For example, it is possible to change the chemical potential $\mu$ and the interaction strength $\alpha$ of the proteins through RNAi (RNA interference) protocols. It is also possible to change the diffusivity $D_p$ of the proteins through binding with another protein that does not interfere with the interaction between the proteins and the diffusing pronogen. However, $D_p$ only sets the time scale for the proteins to reach their steady state without affecting the steady-state gradient. $\mu$ and $\alpha$, on the other hand, have strong effects on the steady-state protein gradient. The parameter $\bb$, controlling the gradients is
\begin{equation}
 \bb=\f{j_0\bar{A}e^{\mu\alpha}}{R_0+\bar{A}e^\mu}=\f{j_0e^{\mu\alpha}}{R_0/\bar{A}+e^\mu}
\end{equation}
We set $j_0$ and $R_0/\bar{A}$ to unity by rescaling the amplitude at $x=0$ and the units of length and time scales. In Fig. \ref{diffmu_alpha} (a) and (b), we show the effects on the protein gradients when we change $\mu$ from 1 to 2 with $\alpha=0.2$ and for change in $\alpha$ from $0.2$ to $0.4$ with $\mu=1.0$ respectively for the synergistic scenario. The change in the gradient for the antagonistic scenario (negative values of $\alpha$) for the corresponding changes in the parameters are shown in Fig. \ref{diffmu_alpha} (c) and (d). We find that changing $\alpha$ has a stronger effect on the protein gradient compared with changes in $\mu$.

In the cases of interest here, the proteins are significantly larger and thus diffuse more slowly than the pronogens. Thus, the pronogens can reach their steady-state profile adiabatically with respect to the local and slowly evolving, instantaneous protein concentration. That means that the steady-state pronogen profile is a function of the local protein concentration $c(x,t)$ which varies only slowly with time, so that on relatively short time scales compared to the protein diffusional dynamics, the pronogens attain their steady state which depends parametrically on $c(x,t)$. Before the pronogens are secreted, the proteins are uniformly distributed in the cell volume and diffuse only locally to adjust their concentration to that of the pronogen. We focus on time scales that are long compared with both the pronogen and protein diffusional time scales and determine the local protein concentration in equilibrium. 
Since the pronogens degrade, their concentration profile is determined by a steady-state balance of the incoming flux and the degradation (that also depends on the local protein concentration) and not by pronogen equilibrium.
At very long times, one therefore solves the steady-state profiles for the pronogen as coupled to the equilibrium protein concentration profile. 
In the particular case of MEX-5 gradient formation in one-cell stage of {\it C. elegans}, the MEX-5 protein concentration starts from a uniform concentration at pronuclear stage and evolves to a stable protein gradient once nuclear envelope breakdown (NEBD) occurs. 
Our predictions for the coupled pronogen-protein profiles apply at this final stage of NEBD \cite{griffin2011}.

Diffusion-degradation models with non-linear degradation terms appear in many different systems in biological physics, and their quantitative analysis is therefore important. In Ref. \cite{dasbiswas2016} a somewhat similar model, that corresponds to the synergistic scenario in this work, was suggested when the long-range nature of mechanical contractility within a cell assembly was coupled to the pronogen degradation (or trapping within the cytoskeleton). In that case, the sign of the nonlinear degradation term depends on the nature of the pronogen cytoskeletal mechanical coupling and, with the appropriate coupling, can also correspond to the antagonistic scenario studied above \cite{kinjal2017}. 
While models which focus upon the robustness and system size scaling of protein gradient \cite{eldar2002,eldar2003,benzvi2008,benzvi2009,benzvi2011,dasbiswas2016} require that the local, biochemical degradation rate $R_0$ (in the linear degradation term in Eq. (\ref{model2_eq26}) where $A=R_0+\bar{A}e^\mu$) be small, our suggestion for protein gradients driven by pronogen diffusion have no such crucial condition.
When $R_0$ dominates over the protein chemical potential contribution to the linear degradation coefficient, $A$, we predict an exponential profile for the diffusing species and the protein gradient follows from Eq. \ref{cxc0} as long as $\alpha$ remains non-zero.
The stability analysis that we have detailed in the SM also tells us about any possibility of pattern formation within the mechanogen model. We have seen that the pronogen profile is completely stable for synergistic interactions. For antagonistic interactions, there are two possible solutions one of which is stable while the other is unstable. When perturbed, the unstable solution dynamically evolves to the stable one. This shows that the basic version of the mechanogen model will not be unstable to pattern formation, which apparently requires additional interactions of degrees of freedom.

In an interesting follow-up to the studies of robustness and scaling of protein gradients \cite{eldar2002,eldar2003,benzvi2008,benzvi2009,benzvi2011}, England and Cardy \cite{england2005} have investigated the effect of time-dependent noise, allowing for fluctuations in the flux, on the profile of a pronogen gradient in a developing embryo. They found that strong enough noise can completely destroy the robustness of the system. 
Applying this to our theory, we note that their \underline{non-linear} stability analysis applies when the the linear degradation term represented by $A$ is equal to zero, the case they considered.  In that situation, the profile can be non-linearly unstable to large enough noise.  Otherwise, the stability analysis we present here applies.  Note that we focus on the coupling of the protein and pronogen profiles which exists whether or not the pronogen profile obeys system-size scaling, which only occurs when $A=0$.

There can be a number of different scenarios by which the pronogen antagonizes the protein. For example, association of the pronogen molecule may increase the total free energy of the protein. Association of the pronogen with the protein may release other molecules associated with that protein. A similar scenario happens in the one-cell-stage {\it C. elegans} embryo where phosphorylation of the MEX-5, when it is associated with mRNA, may dissociate the RNA molecule. In this case, 
the quadratic term in the antagonistic scenario -- that effectively results in release of additional pronogen to the solution -- cannot be larger than the linear degradation term, since only via degradation can the pronogens be removed from solution to associate with the proteins. 
In our theory, this is reflected in the finding that $\bb$ must be less than $1/\sqrt{3}$ if steady-state, pronogen and protein gradients are to be maintained.

Our model predicts protein gradient formation via the simplest possible non-linear term (quadratic in the concentration) in the effective diffusion-degradation equation for the pronogen.
We have presented our theory for irreversible protein-pronogen interaction. The relative irreversibility is connected to the relation of the various rate constants. For large values of $R_0$, the protein-independent rate of pronogen degradation, [or in the absence of protein-pronogen interaction when $\bar{\Lambda}==0$ (identically equals 0)], the term $\bar{\Lambda}[c(x,t),\psi(x,t)]$ in Eq. (\ref{phosphoeq1}) is negligible compared to the local degradation term.
In that case the pronogen assumes an exponential profile; such a pronogen profile is not robust to source fluctuations \cite{eldar2002,eldar2003}. On the other hand, the self-enhanced degradation mechanism, provided by the protein-pronogen interaction here, leads to power-law profiles that are robust (SM) \cite{eldar2002,eldar2003}.
Even if the pronogen gradient is very large, the resulting protein gradient that depends on the value of the coupling $\alpha$ (Eq. \ref{cxc0}), may not be as large, depending on the value of $\alpha$. 

Ref. \cite{griffin2011} shows that a cortical phosphorylation (localized source of phosphorylation at the boundary of the cell) alone is not capable of capturing the experimental results for the gradient formation of MEX-5 proteins and one must look at the cytoplasmic distribution of the phosphate molecules. 
Our theory allows for a cytoplasmic pronogen steady-state profile (e.g., the phosphate molecules in {\it C. elgans}, see Refs. \cite{griffin2011,boyd1996}) as well as the cytoplasmic protein gradient; this offers a more complete scenario even for situations where the differential-diffusion mechanism may apply. 
The scenario focused upon in our theory is independent of the change of the protein diffusivity upon interaction, which is the key feature of differential diffusion models \cite{lipkow2008,daniels2009}. It is possible that the change in free energy also modifies the diffusivity, but this can occur in other ways as well. The ideal systems to test our ideas may be found in systems with intracellular protein gradients where association with the pronogen changes the protein diffusivity only negligibly. 
Of course, in any given system, it is possible that the pronogen and the differential-diffusion mechanisms both operate. However, the pronogen scenario we have treated, allows cells to control their protein gradients by localized secretion of the morphogen-like molecules.

\section{Supplementary Material}
The supplementary material, available at (url), contains a detailed mathematical analysis of the differential diffusion model showing why the mere presence of slow and fast diffusive species is not enough to establish the validity of the model and motivates the needs for alternative mechanisms, an outline of the conditions where the assumption of equilibrium consideration of the protein profile is valid, the stability analyses for the two cases and a discussion of the stability analysis in terms of an energy functional.

\acknowledgements{We would like to thank Kinjal Dasbiswas, Ohad Cohen, David J. Odde, Nir S. Gov, S. Saha and E. Griffin for many important discussions and suggestions. SKN thanks Koshland foundation for funding through a fellowship. 
SAS is grateful for the support of the Israel Science Foundation and the US-Israel Binational Science Foundation as well as continuing support from the Perlman Family Foundation.}


%

 \vspace{1cm}
 
 \section*{Supplementary Material}
 
\twocolumngrid
\renewcommand{\theequation}{S\arabic{equation}}
\renewcommand{\thefigure}{S\arabic{figure}}
\renewcommand{\thesection}{ S\arabic{section}}

\setcounter{equation}{0}
\setcounter{figure}{0}
\setcounter{section}{0}

This supplementary material contains a detailed mathematical analysis of the differential diffusion model showing why the mere presence of slow and fast diffusive species is not enough to establish the validity of the model and motivates the needs for alternative mechanisms, an outline of the conditions where the assumption of equilibrium consideration of the protein profile is valid, the stability analyses for the synergistic and antagonistic cases and a discussion of the stability analysis in terms of an energy functional.

\section{Differential Diffusion Model: Motivation for alternative mechanism}
\label{motivation_altmech}
\begin{figure}
 \includegraphics[width=8.6cm]{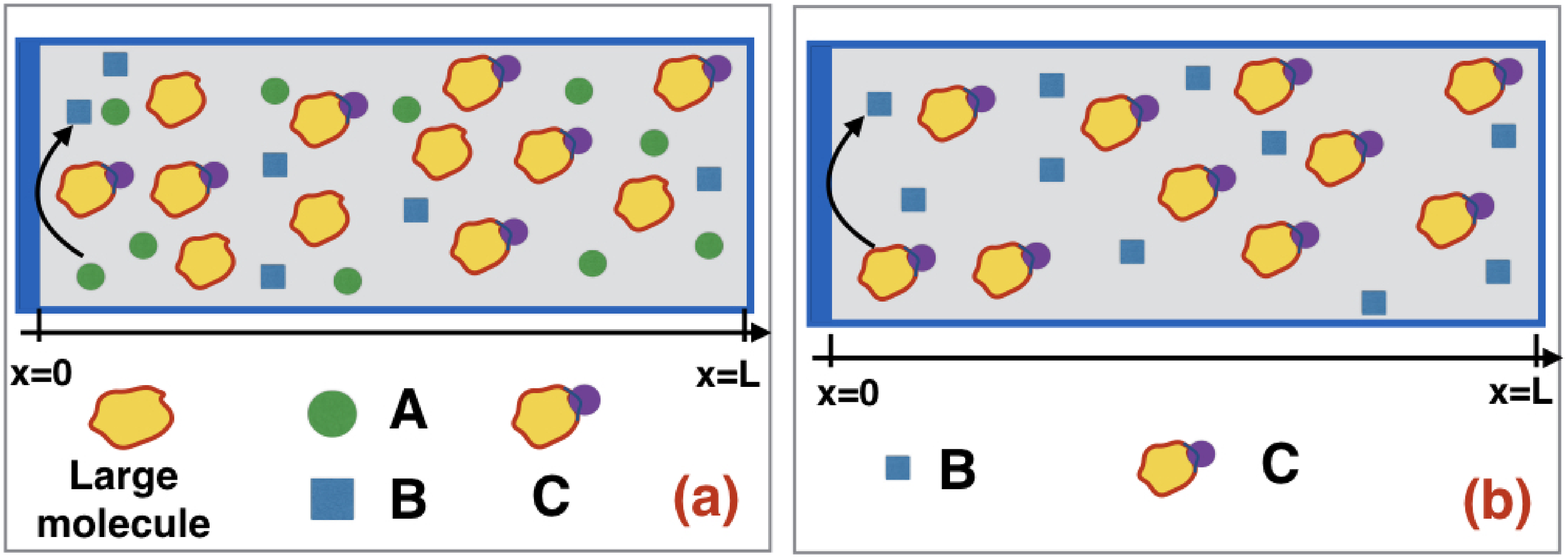}
 \caption{We consider two limiting cases of the differential diffusion model in the context of MEX-5 protein gradients within a single cell in {\it C. elegans}. (a) \underline{Case 1}: A system of three states of the same protein where $A$ is the unphosphorylated free form, $B$ is the phosphorylated form and $C$ is the form when the protein is associated with a larger molecule, which can happen throughout the cell volume. $A$ is phosphorylated at the boundary of the cell at $x=0$. $C$ has a very small diffusivity whereas the diffusivities of $A$ and $B$ are similar. Since the diffusivity of $C$ is very small, it is not relevant whether $C$ is modified by phosphorylation since the probability of $C$ reaching the boundary where phosphorylation occurs is very small. The protein dissociates from the larger molecules with a certain rate and only those free MEX-5 proteins reach the boundary and become phosphorylated. 
 We show that gradient formation in such a system cannot be explained by the differential diffusion models. (b) \underline{Case 2}: A system of two states of the same protein: $B$ with large diffusivity and $C$ bound to a large molecule so that the complex has a small diffusivity. When $C$ is phosphorylated at $x=0$, the complex dissociates and $C$ transforms to $B$. $B$ transforms into $C$ throughout the system since dephosphorylation and complex formation can occur stochastically in the entire bulk. Differential diffusion models can explain the gradient formation of the total protein concentration in this case. In both figures, $x=L$ is the system boundary that prevents a flux of proteins outside the cell.}
 \label{phospho_model}
\end{figure}

We first consider the differential diffusion model for protein gradients within a single cell, as proposed in \cite{SMlipkow2008,SMdaniels2009}. This involves three species of the same protein that can be in three different states, $A$, $B$ and $C$ with concentrations $c_A$, $c_B$ and $c_C$ respectively as schematically shown in Fig. \ref{phospho_model}(a). In the gradient formation of MEX-5 protein at the one-cell-stage of {\it C. elegans} embryo the amount of total protein is larger at the anterior side and the protein is phosphorylated at posterior end of the embryo. MEX-5 binds to larger molecules through RNA with a certain rate and in a manner that is uniform throughout the cell volume. When phosphorylated, the rate of binding to RNA (and hence to the larger molecules) is extremely small, we assume it zero for simplicity. We further simplify the picture through the assumption that binding of the dephosphorylated form of MEX-5 to RNA directly implies binding to the larger molecules ignoring the detailed mechanism of this process. 

When bound to larger molecules, the MEX-5 large molecule complex has a very small diffusivity. 
In the absence of the detailed understanding of those binding (and unbinding) mechanism, we consider two limiting cases. 
(1) Since MEX-5 large molecule complex has very small diffusivity, it doesn't reach the posterior cortex (the only region where phosphorylation can occur) and remains dephosphorylated. However, MEX-5 stochastically dissociate from the larger molecule with some rate; this allows MEX-5 to diffuse so that it can reach the phosphorylation site and become phosphorylated as in Fig. \ref{phospho_model}(a). In this case, phosphorylation does not change the diffusivity of the protein but dephosphorylation, which happens uniformly throughout the system allows the proteins to be associated with the larger molecules, and thus, changes the diffusivity. We show below that this scenario does not lead to MEX-5 gradient formation through the mechanism of differential diffusion. 
(2) Phosphorylation affects the complex of MEX-5 bound to the larger molecule and allows the MEX-5 to unbind from the larger molecule. The MEX-5 diffusivity then increases as in Fig. \ref{phospho_model}(b). 
We show that this scenario can lead to MEX-5 gradient formation through the mechanism of differential diffusion as has been shown in \cite{SMgriffin2011}.
An important conclusion of our paper is that although the differential diffusion mechanism does not apply in case (1) our model, presented in Sec. II, predicts a protein gradient even in this case.

We now analyze the differential diffusion model for case (1), demonstrate that it cannot explain gradient formation and then discuss case (2) that can allow for gradient formation.
Within our description, state $A$ is the ordinary form of the protein (MEX-5) that is phosphorylated (though it can be any other molecule as well) at the boundary at $x=0$ with a rate $k_p$, so that it is transformed to state $B$. In state $A$, the protein can associate with the larger molecules, uniformly distributed in the bulk of the cell and we denote the complex as state $C$ that occurs with a rate $k_b$; the dissociation rate of $C$ reverting to the large molecule and the protein in state $A$ is $k_d$. $x=L$ is the cell boundary where we assume no-flux boundary conditions for all species of the proteins and complexes.

\underline{Case 1}: 
Since the proteins are dephosphorylated uniformly throughout the system, we assume $B$ converts to $A$ with a rate $\tilde{k}$ in the bulk of the cell.
It is reasonable to assume that the change in diffusivity of the proteins due to phosphorylation is very small (much smaller than even a factor of 2) and the diffusivities of $A$ and $B$ should be similar. However, we first consider the general case and allow these diffusivities to be different and specialize later on. The diffusivities of $A$, $B$ and $C$ are denoted by $D_A$, $D_B$ and $D_C$ respectively. The set of first order reactions that represent the scenario described here are:
\begin{equation}\label{phosphomodel_reactions}
 A\xrightarrow{k_p} B, \,\,\, A\xrightleftharpoons[k_d]{k_b} C, \,\,\, B\xrightarrow{\tilde{k}}A
\end{equation}
where the first reaction in Eq. (\ref{phosphomodel_reactions}) occurs at the boundary $x=0$ and affects the concentration profiles through boundary conditions, while the rest of the reactions take place in the bulk of the cell. We now have three equations describing the dynamics of the concentrations of the three states of the protein:
\begin{align}\label{phospho_model_timeeqs}
\text{Case 1: } \,\,\, \begin{cases}
 \f{\p c_A}{\p t} &= D_A\f{\p^2c_A}{\p x^2}-k_b c_A+k_dc_C+\tilde{k}c_B \\
 \f{\p c_B}{\p t} &= D_B\f{\p^2 c_B}{\p x^2}-\tilde{k} c_B\\
 \f{\p c_C}{\p t} &= D_C\f{\p^2 c_C}{\p x^2}+k_b c_A-k_dc_C,
 \end{cases}
 \end{align}
with the boundary condition for C being 
\begin{align}
 -D_C\f{\p c_C}{\p x}\bigg|_{x=0}=-D_C\f{\p c_C}{\p x}\bigg|_{x=L}=0
\end{align}
since $C$ is neither created nor degraded at either of the boundaries. The boundary conditions at $x=0$ for $A$ and $B$ contain the results of phosphorylation that occur only there:
\begin{align}\label{ABbc}
  -D_A\f{\p c_A}{\p x}\bigg|_{x=0}=-k_p c_A, \,\,\,-D_A\f{\p c_A}{\p x}\bigg|_{x=L}=0 \nonumber\\
  -D_B\f{\p c_B}{\p x}\bigg|_{x=0}=k_p c_A, \,\,\,-D_B\f{\p c_B}{\p x}\bigg|_{x=L}=0.
\end{align}
We now find the steady-state solution where the time derivatives are all zero. Adding Eqs. (\ref{phospho_model_timeeqs}), and taking a first integral, we obtain
\begin{equation}
 D_A\f{\p c_A}{\p x}+D_B\f{\p c_B}{\p x}+D_C\f{\p c_C}{\p x}=c_1
\end{equation}
where $c_1$ is an integration constant. Using the boundary conditions at $x=0$ or $x=L$, we obtain $c_1=0$. Then,
\begin{equation}\label{phospho_model_argument}
 D_A\f{\p c_A}{\p x}+D_B\f{\p c_B}{\p x}+D_C\f{\p c_C}{\p x}=0.
\end{equation}
We now use the reasonable assumption that $D_A\sim D_B=D$, since phosphorylation cannot result in a significant (e.g., factor of 2) change in the diffusivity of the protein.
When the protein associates with the large molecule, the diffusivity of the complex is much smaller, therefore, $D_c\ll D$. This can be seen in the context of MEX-5 proteins in the one-cell-stage of {\it C. elegans}, $D\sim 20 \mu m^2/s$ whereas $D_C\sim 0.5 \mu m^2/s$ \cite{SMdaniels2010,SMgriffin2011}. \cite{SMdaniels2010} shows that there is a third component with even shower diffusivity $D_C\sim 0.07 \mu m^2/s$. In these limits, we find that ${\p^2(c_A+c_B)}/{\p x^2}=0$. Integrating over $x$ and using the boundary conditions, we find that ${\p (c_A+c_B)}/{\p x}=0$ 
and from Eq. (\ref{phospho_model_argument}) $\p c_C/\p x\approx 0$. Therefore, $\p(c_A+c_B+c_C)/\p x\approx0$ with corrections of the order of the very small difference $(D_A-D_B)$ and $D_C$. Thus, for this case, the
differential diffusion model predicts an approximately zero gradient of the total protein concentration.

\underline{Case 2}: 
For case (2), identical with the models of \cite{SMlipkow2008,SMdaniels2009}, there is no unphosphorylated state that is not bound to the larger molecules, that is state $A$ does not exist. If the protein is phosphorylated, that is, in state $B$, it is not bound to larger molecules. When unphosphorylated, the protein is always bound to the larger molecules, that is, the protein is in state $C$. The protein in state $C$ gets directly phosphorylated and produces $B$ as shown in Fig. \ref{phospho_model}(b). In this case, $D_C\ll D_B$. Thus, the kinetic reactions that represent the scenario are:
\begin{equation}
 C\xrightarrow{k_p}B, \,\,\, B\xrightarrow{k_b}C
\end{equation}
where the first reaction occurs at the boundary at $x=0$ and the second one takes place throughout the system.
The equations describing this case are
\begin{align}\text{Case 2: } \,\,\,
\begin{cases}\label{case2_diffeq}
D_B\f{\p^2c_B}{\p x^2}-{k_b}c_B=0 \\
D_C\f{\p^2 c_C}{\p x^2}+{k_b} c_B=0
\end{cases} 
\end{align}
with the boundary conditions given by
\begin{align}\label{case2_bc}
&  -D_B\f{\p c_B}{\p x}\bigg|_{x=0}=k_p c_C, \,\,\,-D_B\f{\p c_C}{\p x}\bigg|_{x=L}=0 \nonumber\\
&  -D_C\f{\p c_C}{\p x}\bigg|_{x=0}=-k_p c_C, \,\,\,-D_C\f{\p c_C}{\p x}\bigg|_{x=L}=0.
\end{align}
Adding the two equations in Eq. (\ref{case2_diffeq}), integrating once and using the boundary conditions in Eq. (\ref{case2_bc}) we obtain
\begin{equation}
 D_B\f{\p c_B}{\p x}+D_C\f{\p c_C}{\p x}=0.
\end{equation}
Since $D_C\ll D_B$, we find $\p (c_B+c_C)/\p x\neq0$, thus differential diffusion model leads to a gradient of the total protein concentration in this case as has been shown in Refs. \cite{SMlipkow2008,SMdaniels2009,SMgriffin2011}.

Analysis of the two cases above shows that the applicability of the differential diffusion mechanism for the case of {\it C. elegans} strongly depends on the detailed mechanism of how the slow and fast moving species of the proteins arise. The mere presence of the two species in the system does not imply a gradient of the total protein concentration. In the experiments of \cite{SMdaniels2009,SMgriffin2011}, it is found that there are at least two different species of slow and fast moving MEX-5 proteins. Our analysis emphasizes that it is also important to verify if the slow moving species produces the fast moving species at the posterior side upon phosphorylation alone or whether this kinetic process occurs uniformly throughout the system. If it is the latter, the differential diffusion mechanism does not imply a gradient of protein concentration. In any case, our analysis illustrates that the detailed knowledge of the molecular origins of the slow and fast moving species is essential to understand the mechanism of gradient formation of MEX-5. On the other hand, as we show in the main paper, our model that involves the action of a pronogen, a morphogen-like molecule, on the protein is {\it independent} of the diffusivities of the protein states and predicts a non-zero gradient of the total protein concentration for both cases. For the {\it C. elegans} embryo, it is possible that the actual process is a combination of the two limiting cases we considered.

\section{Justification for the equilibrium consideration of the protein profile}
We show in this appendix that the assumption of non-degradation of proteins leads to an equilibrium for the protein distribution at long times. The free energy for the protein concentration $c(x)$ is
\begin{equation}
 f[c(x),\psi(x,t)]=k_BTc(\ln c-1)-\alpha \psi(x,t)c(x).
\end{equation}
Then the chemical potential $\mu\equiv \delta f/\delta c=k_BT\ln c-\alpha\psi$. The dynamics of the protein concentration, with diffusivity $D_p$, is obtained as
\begin{equation}
 \f{\p c}{\p t}=D_p\nabla\cdot[c\nabla\mu]=D_p\nabla[k_BT\nabla c-c\alpha\nabla\psi].
\end{equation}
Therefore, in the steady state, we have
\begin{equation}\label{eqconsid}
 k_BT\nabla c-c\alpha\nabla\psi=\text{const}.
\end{equation}
The constant above in the right hand side (RHS) of Eq. (\ref{eqconsid}) is zero since there can not be any gradient of the chemical potential in equilibrium. Considering the Langevin equation for the positional dynamics of a single protein molecule, we are now going to show that this constant is indeed zero for the scenarios of our interest in this work.

The Langevin equation for the position, $X_p(t)$, of a protein molecule at time $t$ is
\begin{equation}
 \gamma \dot{X}_p(x)=\mathcal{F}_p+\xi_p(t)
\end{equation}
where $\gamma$ is a friction coefficient, $\mathcal{F}_p=\nabla(\alpha\psi)$ is the force on the molecule and $\xi_p(t)$ is Gaussian white noise. The Fokker-Planck equation corresponding to this Langevin equation is obtained \cite{SMriskenbook} as 
\begin{equation}
 \f{\p c}{\p t}=\gamma\f{\p}{\p X_p}[\alpha(\nabla\psi) c]+D_p\f{\p^2c}{\p X_p^2}.
\end{equation}
In equilibrium we have $\p c/\p t=0$ and after integrating the above equation obtain
\begin{equation}\label{eqdist}
 c(x)=c_0e^{-\alpha\psi(x,t\to\infty)/k_BT},
\end{equation}
where we have used $D/\gamma=k_BT$ and $c_0$ is the concentration at $\infty$ when $\psi=0$. Using Eq. (\ref{eqdist}) in Eq. (\ref{eqconsid}) we find that the constant in the RHS is zero.
This justifies our approach of considering the equilibrium for the protein profile at long time while we consider the nonequilibrium dynamics for the pronogen profile.

\section{Stability analysis for synergistic scenario}
The steady-state pronogen concentration, $\phi(x)$, scaled with the flux $j_0$ is $\phi_0(x)=\phi(x)/j_0$. After redefining $x\to\sqrt{A} x$, $t\to At$ and $\bb=j_0B/A$, we obtain the equation of motion for $\phi_0$ as
\begin{equation}\label{model1eq17}
 \f{\d^2\phi_0(x)}{\d x^2}-\phi_0-\bar{B}\phi_0^2=0.
\end{equation}
Multiplying Eq. (\ref{model1eq17}) with $\d\phi_0/\d x$ and integrating once, we obtain
\begin{equation}
 \f{1}{2}\left[\f{\d\phi_0}{\d x}\right]^2-\f{1}{2}\phi_0^2-\f{\bar{B}}{3}\phi_0^3=c_1
\end{equation}
where $c_1$ is the integration constant. As $x\to\infty$, both $\phi_0$ and its derivative must vanish so that we find $c_1=0$. We then write:
\begin{equation}\label{firstint_syng}
 \f{\d\phi_0}{\d x}=\pm \phi_0\left(1+\f{2\bb}{3}\phi_0\right)^{1/2}.
\end{equation}

We now examine the stability of the solution of Eq. (\ref{firstint_syng}). 
For this we perform a linear stability analysis of $\psi(x,t)=\phi_0(x)+\dpsi(x,t)$, where $\dpsi(x,t)$ is a small perturbation to $\phi_0(x)$. Assuming that $\delta\psi$ is small, and keeping terms up to linear order, we obtain from Eq. (9) in the main text,
\begin{equation}\label{sol_model1_eq22}
 \f{\p\dpsi(x,t)}{\p t}=\f{\p^2\dpsi(x,t)}{\p x^2}-\dpsi(x,t)-2\bb\phi_0\dpsi(x,t),
\end{equation}
with the boundary conditions of no-flux for $\dpsi(x,t)$ at both boundaries since the flux boundary conditions are already satisfied the steady-state solution $\phi_0(x)$.
We write $\dpsi(x,t)=\bdpsi(x)e^{\o t}$; as discussed in Sec. \ref{omegafromenergy}, a positive value of $\o$ signifies that the steady state solution is unstable while negative $\o$ indicates a stable solution. Then, from Eq. (\ref{sol_model1_eq22}), we obtain
\begin{equation}
 \f{\p^2\bdpsi(x)}{\p x^2}-(1+\o)\bdpsi(x)-3\csch^2\left(\f{x}{2}+k\right)\bdpsi(x)=0,
\end{equation}
where $k$ is the integration constant obtained from Eq. (13). We see that there is only one real solution of $k$ and two imaginary solutions for a particular value of $\bb$. The imaginary values of $k$ leads to negative values of $\phi_0(x)$; since $\phi_0(x)$ is a concentration and must be positive, we reject these values of $k$ and use the real value only.
We define $\bar{x}=x/2+k$ and write the above equation as
\begin{equation}
 \f{\p^2\bdpsi(\bar{x})}{\p \bar{x}^2}-4(1+\o)\bdpsi(\bar{x})-12\csch^2\bar{x}\bdpsi(\bar{x})=0.
\end{equation}
Defining $x'=\coth \bar{x}$, we obtain from the above equation:
\begin{align}\label{sol_model1_eq25}
 (1-x'^2)&\f{\p^2\bdpsi({x'})}{\p {x'}^2}-2x'\f{\p\bdpsi(x')}{\p x'} \nonumber\\
 +&\left[\nu(\nu+1)-\f{\mu^2}{1-x'^2}\right]\bdpsi(x')=0
\end{align}
with $\nu=3$ and $\mu^2=4(1+\o)$. Eq. (\ref{sol_model1_eq25}) is the standard equation for the associated Legendre function and obtain the solutions in terms of those functions, $P_\nu^\mu(x')$ and $Q_\nu^\mu(x')$ \cite{SMmathewswalker}. Satisfying the no-flux boundary condition at $x=0$, we obtain the value for $\mu$. We find no real solution for $\mu$, but there are imaginary values of $\mu$ that satisfy the boundary condition. This means $\o$ must be negative and that the steady state solution, Eq. (12) is stable. We have verified these analytical results via numerical calculations.

\section{Stability analysis for antagonistic scenario}
Using the scaled variables, as discussed in Sec. IV, we obtain the equation for the steady-state pronogen concentration as
\begin{equation}\label{model2_eq27}
 \f{\d^2\phi_0(x)}{\d x^2}-\phi_0(x)+\bb\phi_0^2(x)=0
\end{equation}
where $\bb=j_0B/A$ and $\phi_0(x)=\phi(x)/j_0$ with $j_0$ is the pronogen flux at its source at $x=0$. Multiplying the above equation by $\d\phi_0(x)/\d x$ and integrating yields
\begin{equation}
 \f{1}{2}\left[\f{\d\phi_0}{\d x}\right]^2-\f{1}{2}\phi_0^2+\f{\bar{B}}{3}\phi_0^3=c_1
\end{equation}
where $c_1$ is the integration constant. Both $\phi_0$ and its derivative must vanish at $x\to\infty$ so that we obtain $c_1=0$. Then,
\begin{equation}
 \f{\d\phi_0}{\d x}=\pm \phi_0\left(1-\f{2\bb}{3}\phi_0\right)^{1/2}.
\end{equation}
Both positive and negative signs lead to the same solution which we find is:
\begin{equation}\label{sssol_antag}
 \phi_0(x)=\f{3}{2\bb}\sech^2\left[\f{x}{2}+k\right],
\end{equation}
where $k$ is the integration constant, as discussed in Sec. IV in the main text.

To examine the stability of the steady-state solution, we analyze the effects of perturbations to $\phi_0$. We write $\psi(x,t)=\phi_0(x)+\dpsi(x,t)$ where we assume that $\dpsi(x,t)$ is small. Then, from Eq. (14), up to linear order in $\dpsi(x,t)$, we obtain
\begin{equation}\label{model2_eq32}
  \f{\p\dpsi(x,t)}{\p t}=\f{\p^2\dpsi(x,t)}{\p x^2}-\dpsi(x,t)+2\bb\phi_0\dpsi(x,t)
\end{equation}
with no-flux of $\dpsi(x,t)$ (i.e., $\d\dpsi/\d x=0$) at $x=0$, since the flux boundary condition has already been satisfied by the steady state soluiton , $\phi_0$.  Similarly, $\delta \psi$ must vanish as $x \to \infty$. 
For the stability analysis we write $\dpsi(x,t)=\bdpsi(x)e^{\o t}$ and find the value of $\omega$ for which the solution for $\delta \psi (x, t)$ satisfies the boundary conditions. 
Positive $\o$ means the solution is unstable and vice versa (see Sec. \ref{omegafromenergy}). We then write:
\begin{equation}
  \f{\p^2\bdpsi(x)}{\p x^2}-(1+\o)\bdpsi(x)+3\sech^2\left(\f{x}{2}+k\right)\bdpsi(x)=0.
\end{equation}
Defining $\bar{x}=x/2+k$ we obtain from the above equation
\begin{equation}
 \f{\p^2\bdpsi(\bar{x})}{\p \bar{x}^2}-4(1+\o)\bdpsi(\bar{x})+12\sech^2\bar{x}\bdpsi(\bar{x})=0.
\end{equation}
We define $x'=\tanh \bar{x}$, and obtain
\begin{align}\label{sol_model2_eq35}
 (1-x'^2)&\f{\p^2\bdpsi({x'})}{\p {x'}^2}-2x'\f{\p\bdpsi(x')}{\p x'} \nonumber\\
 +&\left[\nu(\nu+1)-\f{\mu^2}{1-x'^2}\right]\bdpsi(x')=0
\end{align}
where $\nu=3$ and $\mu^2=4(1+\o)$. Again, this is the standard associated Legendre equation whose solutions are: $P_\nu^\mu(x')$ and $Q_\nu^{\mu}(x')$ \cite{SMmathewswalker}. 
The general solution is the linear combination of these two functions:
\begin{equation}\label{model2_eq36}
 \bdpsi(x')=P_\nu^\mu(x')+DQ_\nu^\mu(x')
\end{equation}
where we have set the overall amplitude of the perturbation to unity (so that $D$ is the ratio of the two Legendre functions), since the overall amplitude is not determined by the boundary conditions.
We have seen that for $\mu\geq 3$ Eq. (\ref{sol_model2_eq35}) has no soluiton that satisfy the boundary condition of no flux at $x=0$. We note that $x'\to 1$ as $x\to\infty$ and $x'=\tanh k$ when $x=0$. To determine the constant $D$ in Eq. (\ref{model2_eq36}) we use the boundary condition $\bdpsi(x')\to 0$ as $x'\to 1^{-}$ (approaching from left where $x<1$). We first obtain the expansions for the associated Legendre functions for $x'\to1^-$ as
\begin{align}
 P_\nu^\mu(x') &\rightarrow \f{1}{\Gamma(1-\mu)}\left(\f{2}{1-x'}\right)^{\mu/2} \\
 Q_\nu^\mu(x') &\rightarrow \f{1}{2}\cos(\mu\pi)\Gamma(\mu)\left(\f{2}{1-x'}\right)^{\mu/2},
\end{align}
and obtain $D$ as
\begin{equation}
 D=-\f{2}{\Gamma(\mu)\Gamma(1-\mu)\cos(\mu\pi)}.
\end{equation}
Note that this result is applicable for non-integer $\mu$. 

For each value of $\bb$ we have two values of $k$ [see Eq. (16)]. For each value of $k$, we obtain a solution for $\mu$ by requiring that the perturbation satisfy the no-flux boundary condition at $x=0$. If the value of $\mu$ is larger than $2$, we find that the eigenvalue $\o$ is positive since $4(1+\o)=\mu^2$, so that the corresponding solution is unstable. On the other hand, if $\mu<2$, then $\o$ is negative and the corresponding solution is stable. We find that for fixed values of $\bb<1/\sqrt{3}$, one of the two solutions corresponding to the two real values of $k$ is stable and the other one is unstable and dynamically evolves to the stable solution.

\begin{figure}
 \includegraphics[width=8.6cm]{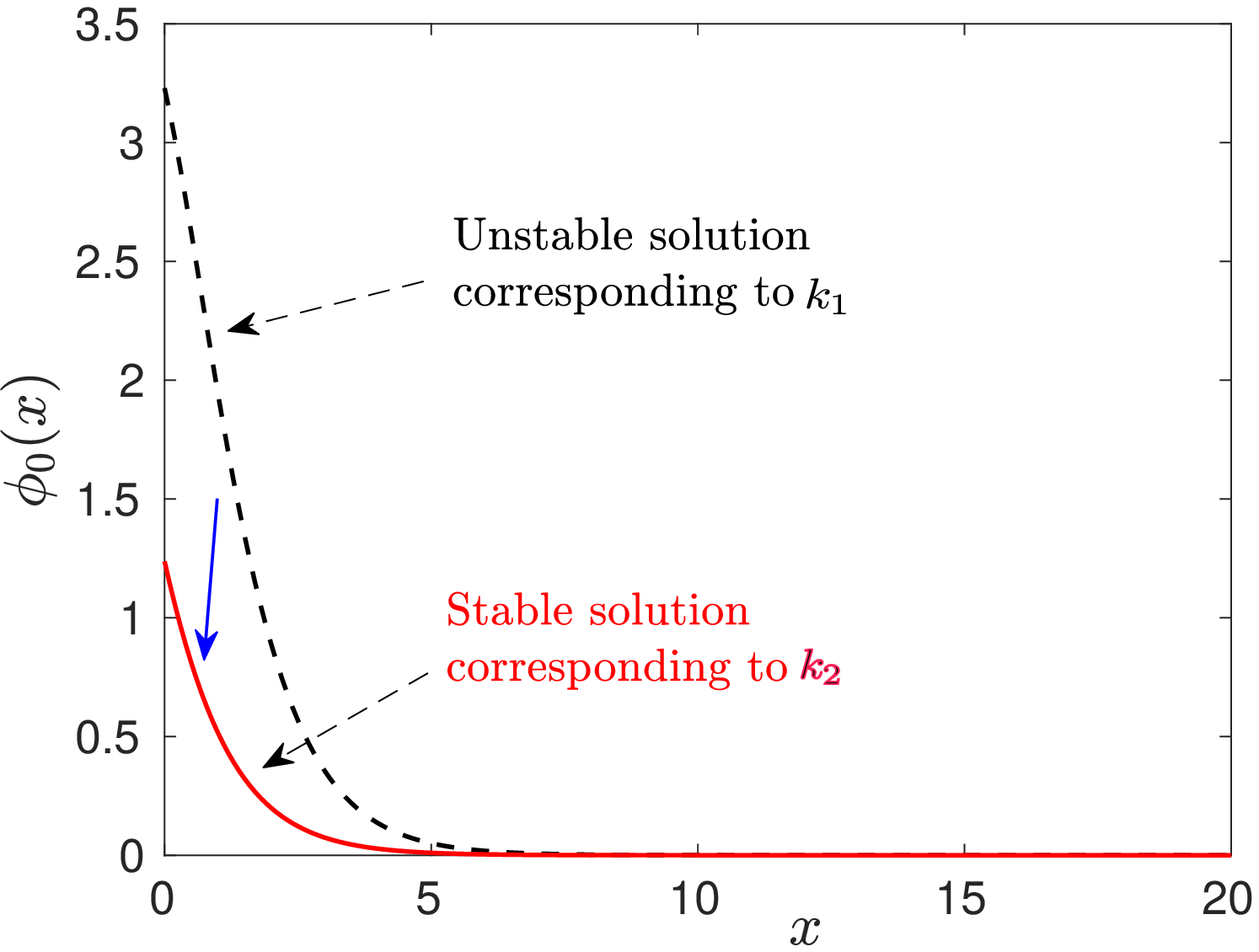}
 \caption{There are two possible solutions for the antagonistic scenario one of which is stable and the other one is unstable. The unstable solution dynamically evolves to the stable solution (see text for more details).}
 \label{model2_bb_0.42}
\end{figure}

We illustrate this point with an example. $\bb$ must be less than $1/\sqrt{3}$ for a physical solution and we choose $\bb=0.42$. We numerically solve Eq. (14) with $\psi(x,t)=\phi_0(x)+G\dpsi(x,t)$ where $G$ is a small amplitude of the perturbation. For the numerical solution, we take $0\leq x\leq 20$ and $0\leq t\leq 400$ and start with the initial condition given by the two solutions for $\phi_0(x)$ and $\dpsi(x,t)$, Eqs. (\ref{sssol_antag}) and (\ref{model2_eq36}) respectively. For $\bb=0.42$, we obtain the two values of $k$ as $k_1=0.32$ and $k_2=1.12$. For $k_1$ we obtain $\mu=2.58$ and for $k_2$ we obtain $\mu=1.22$. Therefore, from our analysis, the solution corresponding to $k_2$ is stable whereas the solution corresponding to $k_1$ is unstable and evolves to the solution corresponding to $k_2$ with time. This is exactly what we observe in the numerical solution as illustrated in Fig. \ref{model2_bb_0.42}.

What are the differences between the two solutions when their shapes look so similar, while one is stable and the other unstable? For the example that we have considered here, if we integrate both solutions over $x$ we find a total scaled amount of pronogen equal to $4.93$ for the solution corresponding to $k_1$ and $1.37$ for the solution corresponding to $k_2$. Therefore, we see that when the amount of pronogen is large, the solution becomes unstable and dynamically evolves to the one that corresponds to smaller amount of pronogen in steady-state. We can understand this stability in another way. Let us look at the last two terms of Eq. (\ref{model2_eq27}) where the first one reduces the rate of pronogen accumulation, while the last one increases this rate.
$\phi_0(x)$ has the maximum at $x=0$. If $\bb\phi_0(x=0)$ is greater than unity, the accumulation term dominates the degradation term and the concentration profile is unstable; otherwise, it is stable. For the example we have considered, we obtain $\bb\phi_0(x=0)=1.36$ for the solution corresponding to $k_1$ and $\bb\phi_0(x=0)=0.52$ for the solution corresponding to $k_2$. Thus, the former solution is unstable and the latter is stable, similar to what we conclude from more detailed numerical and analytical analysis.


\section{Stability analysis in terms of variation of a functional}
\label{omegafromenergy}
For concreteness, we consider the antagonistic scenario where the equation of motion for the pronogen density is given by
\begin{equation}\label{appendixeq1}
\f{\p\psi(x,t)}{\p t}=D\nabla^2\psi(x,t)- A\psi(x,t) + B\psi^2(x,t).
\end{equation}
It is well known \cite{SMhohenberghalperin} that systems near thermodynamic equilibrium have dynamics that can often be related to variations of their free energy.
Though the biological systems we consider here are, by their nature, far from equilibrium, in the cases treated here, we can still define a functional of the pronogen concentration, $\mathcal{F}$ whose variation governs the dynamics. This functional is \underline{not} to be associated with any free energy since it is determined by various reaction processes. We thus write:
\begin{equation}\label{appendixeq2}
 \mathcal{F}=\int\d x \left[\f{D}{2}\left(\f{\p \psi(x,t)}{\p x}\right)^2+\f{A}{2}\psi^2(x,t)-\f{B}{3}\psi^3(x,t)\right]
\end{equation}
and the dynamics is obtained from the equation
\begin{equation}\label{appendixeq3}
 \f{\p\psi(x,t)}{\p t}=-\f{\delta\mathcal{F}}{\delta \psi}.
\end{equation}
It is instructive to understand the stability analysis, by considering Eq. (\ref{appendixeq2}) along with Eq. (\ref{appendixeq3}) that gives rise to the dynamical equation that governs the pronogen.

We now consider fluctuations away from the steady-state pronogen concentration profile and write $\psi(x,t)=\phi_0(x)+\dpsi(x,t)$ where $\dpsi(x,t)$ is small and expand $\mathcal{F}$ up to quadratic order in $\dpsi(x,t)$: 
\begin{align}
 \delta\mathcal{F}&=\int\d x\left[D\f{\p\phi_0}{\p x}\f{\p \dpsi}{\p x}+A\phi_0\dpsi-B\phi_0^2\dpsi\right] \nonumber\\
 &+\int\d x\left[\f{D}{2}\left(\f{\p\dpsi}{\p x}\right)^2+\f{A}{2}\dpsi^2-B\phi_0\dpsi^2\right]
\end{align}
where $\delta\mathcal{F}=\mathcal{F}[\psi]-\mathcal{F}[\phi_0]$. Integrating by parts, and assuming that the perturbation vanishes at the boundaries, we obtain
\begin{align}\label{appendixeq6}
 \delta\mathcal{F}&=\int\d x\left[-D\f{\p^2\phi_0}{\p x^2}+A\phi_0-B\phi_0^2\right]\dpsi \nonumber\\
 &+\int\d x \dpsi\left[-\f{D}{2}\f{\p^2\dpsi}{\p x^2}+\f{A}{2}\dpsi-B\phi_0\dpsi\right].
\end{align}
The first part of the integrand is the equation whose zero determines the steady-state concentration profile $\phi_0(x)$ of Eq. (\ref{appendixeq1}); so this term vanishes.
If we write $\dpsi(x,t)=\bdpsi(x)e^{\o t}$, the second part can be identified, with the help of Eq. (\ref{model2_eq32}), as $-\o\dpsi(x,t)/2$. Then we obtain
\begin{equation}
 \delta\mathcal{F}=-\o\int \d x \f{1}{2}\dpsi^2(x,t).
\end{equation}
This equation shows that if $\o$ is positive, any perturbation decreases the functional and the solution is unstable. On the other hand, if $\o$ is negative, the perturbation $\dpsi$ increases the functional, therefore the system rejects the perturbation and the steady state solution is stable.


%


\end{document}